\documentclass[aps,prb,preprint,linenumbers,groupedaddress,showkeys,
showpacs,amsmath,amsfonts,amssymb]{revtex4-1}

\usepackage{graphicx}
\usepackage{dcolumn}
\usepackage{bm}
\usepackage{gensymb}
\usepackage{epstopdf}
\usepackage[T1]{fontenc} 
\usepackage{lmodern} 

\usepackage{color} 
\newcommand{\tensorss}[1]{{\textit{\textsf{\textbf{#1}}}}}

\begin{document}


\title{Dangling bonds in a-Si:H revisited: A combined Multifrequency EPR and DFT Study}

\author{M.~Fehr}
\email{matthias.fehr@helmholtz-berlin.de}
 \author{A.~Schnegg}
  \author{B.~Rech}
   \author{K.~Lips}%
\affiliation{%
Helmholtz-Zentrum Berlin f\"ur Materialien und Energie, Institut f\"ur Silizium-Photovoltaik, Kekul\'estr.~5, 12489 Berlin, Germany}%

\author{O.~Astakhov}%
	\author{F.~Finger}%
\affiliation{%
Forschungszentrum J\"ulich, Institut f\"ur Energie- und Klimaforschung, Photovoltaik, 52425 J\"ulich, Germany}%

\author{G.~Pfanner}%
	\author{C.~Freysoldt}%
		\author{J.~Neugebauer}%
\affiliation{%
Max-Planck-Institut f\"ur Eisenforschung GmbH, Max-Planck Strasse 1, 40237 D\"usseldorf, Germany}%

\author{R.~Bittl}%
	\author{C.~Teutloff}%
	\email{christian.teutloff@fu-berlin.de}
\affiliation{%
Freie Universit\"at Berlin, Fachbereich Physik, Arnimallee 14, 14195 Berlin, Germany}%

\date{\today}

\begin{abstract}
Multifrequency pulsed electron paramagnetic resonance (EPR) spectroscopy using S-, X-, Q- and W-Band frequencies (3.6, 9.7, 34, and 94 GHz, respectively) was employed to study paramagnetic coordination defects in undoped hydrogenated amorphous silicon (a-Si:H). The improved spectral resolution at high magnetic field reveals a rhombic splitting of the g-tensor with the following principal values: $g_x=2.0079$, $g_y=2.0061$ and $g_z=2.0034$ and shows pronounced g-strain, i.e., the principal values are widely distributed. The multifrequency approach furthermore yields precise $^{29}$Si hyperfine data. Density functional theory (DFT) calculations on 26 computer-generated a-Si:H dangling-bond models yielded g-values close to the experimental data but deviating hyperfine interaction values. We show that paramagnetic coordination defects in a-Si:H are more delocalized than computer-generated dangling-bond defects and discuss models to explain this discrepancy. 
\end{abstract}



\maketitle
\section{Introduction}
The performance of thin-film solar cells and other devices based on hydrogenated amorphous silicon (a-Si:H) is limited by localized defect states in the mobility gap, which act as recombination centers for excess charge carriers. In undoped a-Si:H, the defect centers are often paramagnetic and give rise to an inhomogeneously broadened asymmetric EPR line at around $g=2.0050-2.0055$ \cite{stutzmann1989_prb}. The intensity of this signal is routinely used as a measure for the electronic quality of a-Si:H \cite{stutzmann1989_philmag60}. The impact of these defect centers on the efficiency of solar cells is even aggravated by the fact that the defect density significantly increases upon light exposure \cite{dersch1981_apl}. This light-induced degradation phenomenon, known as the Staebler-Wronski effect (SWE) \cite{staebler1977_apl,staebler1980_jap}, significantly limits the maximum efficiency of solar cells based on a-Si:H \cite{fritzsche2001_annual_rev}. In order to reduce the impact of a-Si:H defects in the degraded state optimized deposition protocols have been developed \cite{zeman2009_philmag}. Despite these improvements, a nanoscopic understanding of the processes leading to the creation of light-induced defects is still missing \cite{fritzsche2001_annual_rev}. Detailed knowledge about the microscopic origin of SWE defects leading to strategies to eliminate them is therefore mandatory to reach ultimate device performance. Moreover, knowledge of the detailed structure of the defect and the distribution of H atoms in its vicinity is of main importance for the models for the SWE. In view of the latest EPR experiments it became evident that in as grown materials the H is randomly distributed with respect to the defect center~\cite{fehr2009_aSi_pss}. If hydrogen is mediating the SWE effect, it is important to find the precursor of such a state.\\
There is a general consensus in the research community that the dominating defects in a-Si:H are intrinsic coordination defects, i.e. over- (fivefold) or undercoordinated (threefold) Si atoms. The latter are usually denoted by dangling-bond (DB) defects.
To conclude on the defect structure, EPR techniques are most valuable, since the EPR spectrum reflects the electronic structure of the paramagnetic defect. In the present case, the EPR spectrum is determined by two interactions, the Zeeman interaction given by the g-tensor, and the hyperfine interaction (HFI) between the unpaired electron spin and nuclear spins of close-by H and Si atoms. While the g-tensor reflects the global electronic defect structure of the paramagnetic defect, HFIs probe the defect wave function locally. Combining these two pieces of information detailed spin-density \cite{footnote1} maps of the unpaired electron spin may be obtained, which constitute highly desired pieces of information to identify the microscopic origin of the defect centers.\\
A detailed analysis of the EPR spectrum of coordination defects was first carried out by Stutzmann \textit{et al.} at a microwave frequency of 9 GHz (X-Band) \cite{stutzmann1989_prb}. They determined the g-tensor of the unpaired electron spin to be axially symmetric with principal values similar to the $P_b$ center occurring at the Si/SiO$_2$ interface \cite{poindexter1981_jap,cook1988_prb}. In a subsequent study, Umeda \textit{et al.}~\cite{umeda1999_prb} revised the g-tensor values by studying the EPR spectrum at different resonance frequencies (S-, X- and Q-band) with increased spectral resolution (see Table \ref{tab:multifreq_study}). However, in both studies the g-tensor was already assumed as axially symmetric in the fitting models and never systematically tested against rhombic symmetry. In addition to the g-tensor, Stutzmann \textit{et al.}~\cite{stutzmann1989_prb} determined the HFI with the nuclear spin of the Si atom where most of the defect spin density is concentrated. The HFI of this particular atom is characterized by an anisotropic tensor, which will be denoted by $\tensorss{A}_\text{L}$ in the following, i.e. the A-tensor with the largest isotropic HFI. By analyzing the principal values of this tensor within an analytical linear combination of atomic orbitals (LCAO) model \cite{watkins1964_pr}, Stutzmann \textit{et al.} \cite{stutzmann1989_prb} determined the wave function of the defect. Since their analysis of the HFI suggests that the defect wave function is sp$^x$-hybridized and of strong p-character, the authors concluded that the electronic structure of the given center resembles a DB similar to the $P_b$ center. However, the isotropic HFI ($A_\text{iso}$) of coordination defects in a-Si:H, as evaluated by Stutzmann and Biegelsen \cite{stutzmann1989_prb,biegelsen1986_prb}, is given by $A_\text{iso} = 200$ MHz and is therefore much smaller than the isotropic HFI of the $P_b$ center ($A_\text{iso} = 315$ MHz) \cite{brower1983_apl,carlos1987_apl}. This discrepancy was attributed to a relaxation of the atomic structure of the DB from a tetrahedral configuration to a more planar geometry, induced by the amorphous environment. It was argued that in the latter configuration the p-character of the DB wave function is enhanced over the s-character, which leads to a smaller isotropic HFI. It is, however, not \textit{a priori} clear whether such a relaxation actually takes place when a DB is created in a-Si:H. This question can only be clarified by a detailed quantitative theoretical treatment of the atomic DB structure and the resulting EPR parameters, which is still missing up to now.\\
The possibility to compare experimental g- and A-tensors in amorphous semiconductor materials with theoretical calculations came into reach only recently. This was mainly due to two reasons. Firstly, precise g-tensor data is usually only available for crystalline materials and secondly, a lack of \textit{ab-initio} approaches capable of calculating g-tensors from complex material structures. This situation changed with the advent of advanced density-functional theory (DFT) methods, which proved to be able to reproduce experimentally determined g-tensors even in complex Si materials \cite{pickard2002_prl,gerstmann2010_pss}. Up to now such studies have been restricted to crystalline Si materials. One of the purposes of this study is to extend this powerful approach to a-Si:H.\\
Here, we present a detailed experimental and theoretical analysis of the g-tensor and the HFIs of the dominant defect center in a-Si:H. We employ high resolution EPR measurements and complement them by DFT calculations capable of relating measured g- and A-tensors to the spin-density distribution, binding geometry and electrostatic surrounding of the paramagnetic site. The defects in a-Si:H are studied experimentally by multifrequency EPR (S-band: 3.6~GHz/0.13~T, X-band: 9.7~GHz/0.34~T, Q-band: 34~GHz/1.2~T and W-band: 94~GHz/3.35~T). In the absence of field-dependent line broadening mechanisms and at high signal to noise ratio (S/N), a high frequency spectrum is generally enough to determine principal values of g- and A-tensors as well as their relative orientation. In the present case, however, pronounced site-to-site variations of the principal g-values (g-strain) restrict the determination of principal A-values at high resonance frequencies. This limitation can be overcome by the multifrequency approach, which allows to separate field-independent ($\tensorss{A}$) and field-dependent ($\tensorss{g}$) spectral contributions. The HFIs are thereby best resolved at low magnetic fields and corresponding frequencies (S-band and X-band), while principal values of the g-tensor can be best determined at high frequencies (Q- and W-band). Furthermore, we applied field-swept echo-detected (FSE) EPR instead of previously used continuous wave (c.w.) EPR since FSE-EPR resolves broad, tailing spectral features better than c.w. EPR techniques. g- and A-values are extracted from experimental FSE spectra by a robust iterative fitting procedure. As a result we show that the g-tensor symmetry of coordination defects in a-Si:H is rhombic and therefore lower than axial symmetry as claimed in earlier studies \cite{stutzmann1989_prb,umeda1999_prb}. This result is important to improve the reliability and precision of g-tensor values and provides the basis for detailed studies of correlations between material properties and $\tensorss{g}$ \cite{astakhov2009_prb,wu1988_prb}. Improved g-tensor values may also help to determine structural differences between light-induced and native defects and thereby shed new light on physical processes underlying the SWE.\\
In order to see whether or not the experimentally determined g- and A-values are adequately described by the atomistic DB model, we employed 26 DB defect structures, generated in two different ways by state of the art annealing techniques~\cite{biswas2009_jpcm,jarolimek2009_prb}. 
Each model contained a single DB defect.
The g- and A-values of each DB model were then calculated by DFT methods
and compared to the experimentally obtained magnetic interaction parameters.

\section{Materials and Methods}
Undoped a-Si:H samples were deposited with plasma-enhanced chemical vapor deposition (PECVD) on a 10 cm $\times$ 10 cm Mo foil at a substrate temperature of about 185~\celsius, of undiluted silane (silane concentration 100~\%), pressure 0.7 mbar, power density of 130 mW/cm$^2$, interelectrode distance of 12 mm, resulting in a deposition rate of 1.8 nm/s. ESR powder samples have been prepared as described in~\cite{xiao2011}. The initial defect density of the samples as determined by c.w. EPR is given by $N_\text{D} = 4(1)\cdot10^{16}$~cm$^{-3}$. The hydrogen content of the sample is about 21 at. \% as determined on a reference a-Si:H sample using Fourier-transform infrared spectroscopy.
The films were removed from the substrate by diluted hydrochloric acid and flakes were collected in EPR-quartz tubes. Pulsed EPR spectroscopy at S-, X- and Q-band was performed on a Bruker BioSpin ElexSys E580 spectrometer and EPR measurements at W-band were performed on an ElexSys E680 spectrometer. The probe heads employed at S-, X- and W-Band were a Bruker ER4118S-MS5, a Bruker ER4118X-MD5, and a Bruker EN600-1021H, respectively. At Q-band a home-built probe head was used. Temperature control was realized with CF935 helium bath cryostats and ITC503 temperature controllers from Oxford Instruments. All experiments were carried out at a temperature of 80 K and utilized a typical field-swept echo (FSE) pulse sequence ($\pi/2-\tau-\pi-\tau-\text{echo}$) with a $\pi$ pulse length of 40 ns, 32 ns, 80 ns, 128 ns and an interpulse delay $\tau$ of 400 ns, 300 ns, 400 ns, 300 ns at S-, X-, Q-, W-band, respectively. Spectra were independent of $\tau$ (data not shown) and the shot-repetition time was set sufficiently long to avoid a saturation of the spin system ($> 2$ ms). EPR spectra were accumulated about 1.5 h at Q- and W-band, and about 10 h at S- and X-band due to lower sensitivity at these frequencies.\\
DFT calculations were carried out with a plane-wave pseudopotential formalism implemented in the Quantum Espresso package \cite{giannozzi2009_jpcm}. We used norm-conserving, scalar-relativistic Troullier-Martins pseudopotentials and the PBE exchange-correlation functional. A plane-wave energy cutoff of 30 Ry ensures convergence with respect to the basis set. The Brillouin zone integration is done on a $6\times6\times6$ Monkhorst-Pack mesh. The HFIs of all atoms in the supercell are determined from a projector augmented wave (PAW)-like post processing step from the self-consistent calculation \cite{vandewalle1993_prb} using two projectors per 1-channel. Please note that the HFIs of $^{29}$Si nuclear spins are mostly negative, since the nuclear g-value of $^{29}$Si ($g_\text{n}$ = -1.1106) is negative. The g-tensor is computed by the GI-PAW formalism \cite{pickard2002_prl}. We consider 26 a-Si:H models consisting of 64 silicon and 8 hydrogen atoms. The defect-free a-Si:H models were created either by releasing hydrogen into Wooten-Winer-Weaire models of a-Si \cite{biswas2009_jpcm}, or by heating and gradually annealing of c-Si:H models followed by structural relaxation \cite{jarolimek2009_prb}. DBs were generated in these models by removing one of the hydrogen atoms, followed by structural relaxation.

\section{Results and Discussion}
Fig.~\ref{fig:multifreq_spectra} depicts FSE spectra of a-Si:H powder samples taken at different microwave frequencies (S-, X-, Q- and W-band, respectively). In the left column (Fig.~\ref{fig:multifreq_spectra} a-d) experimental spectra taken at indicated frequency bands (crosses) are shown together with simulations obtained with parameters given in Table \ref{tab:multifreq_study} (red solid lines).
In the right column (Fig.~\ref{fig:multifreq_spectra} e-h) simulated FSE spectra in the absence of g-, A-strain and broadening due to unresolved HFIs are shown, to make the impact of $\tensorss{g}$ and $\tensorss{A}$ on the line shape at different resonance frequencies more obvious. In the following we will first qualitatively assign the frequency dependence of the EPR spectra to the dominating magnetic interactions. Secondly we describe the fitting routine applied to quantitatively extract the principal g- and A-values. Finally we compare these parameters with values derived from DFT calculations on computed a-Si:H DB models.

\subsection{Analysis of multifrequency EPR spectra}
\begin{figure}
\includegraphics[width=0.6\textwidth]{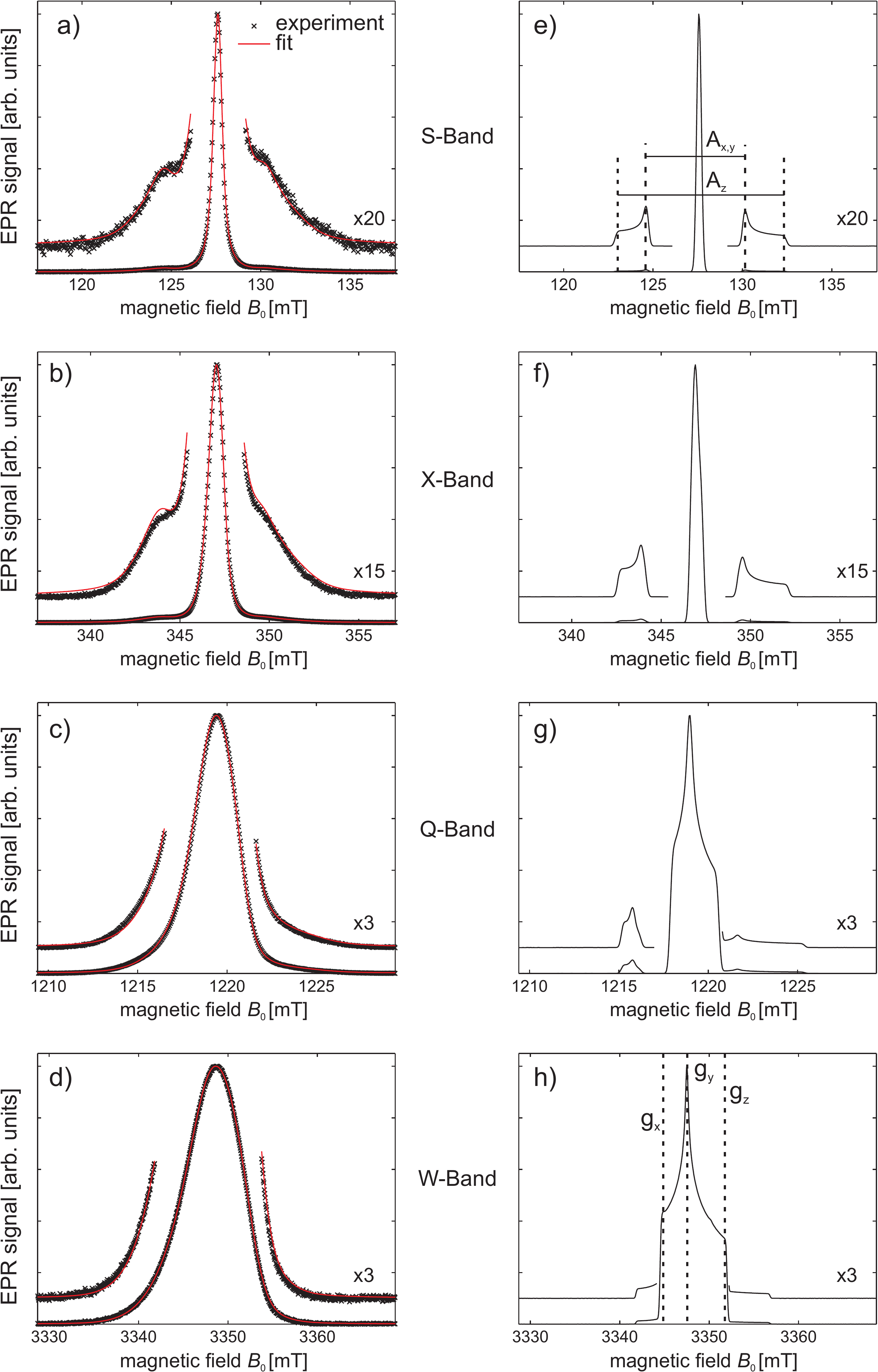}
\caption{S-/X-/Q- and W-band field-swept echo EPR spectra of defects in a-Si:H ($g$ = 2.0055) at a temperature of 80 K. Each spectrum was recorded by integrating the primary echo of a ($\pi/2-\tau-\pi-\tau-\text{echo}$) pulse sequence. Left column (a-d): experimental spectra (crosses) and the fitted spectra (red solid line) obtained with the model described in the text.
Spectra are offset vertically for clarity. Right column (e-h): fitted spectra without g-strain, A-strain and isotropic magnetic field broadening. Principal values of the g-tensor and the A$_\text{L}$-tensor are indicated by the vertical and horizontal lines in e) and h).
\label{fig:multifreq_spectra}}
\end{figure}

The S- and X-band spectra (see Fig.~\ref{fig:multifreq_spectra} a,b) consist of an intense central line and two less intense satellite peaks (see enlarged spectral regions in Fig.~\ref{fig:multifreq_spectra}a,b). Si enrichment studies showed that the EPR spectrum is subject to isotope effects, since naturally abundant Si is composed of stable non-magnetic isotopes ($^{28/30}$Si) with a total abundance of 95.32~\%, and one stable magnetic isotope ($^{29}$Si) with an abundance of
4.68~\% \cite{biegelsen1986_prb}. If the immediate vicinity of the defect is depleted from magnetic isotopes ($^{29}$Si and $^1$H) large HFIs are absent, resulting in the narrow central line, which is broadened by unresolved HFI to more distant $^1$H and $^{29}$Si nuclei \cite{umeda1999_prb,brandt1998_jncs}. This broadening of the resonance line is well described by a Lorentzian in the central part of the line (see Fig.~\ref{fig:multifreq_spectra}a) and its width is proportional to the abundance of $^{29}$Si, $p$, for $p < 10$ \% \cite{umeda1999_prb}. In cases, however, where Si atoms, which exhibit a significant spin density, are magnetic ($^{29}$Si isotope), the EPR spectrum is dominated by large HFIs ($>$ 150 MHz, equivalent to $>$ 7 mT) giving rise to satellite formation in the EPR spectrum. These satellites are symmetrically centered about the narrow central line (see Fig.~\ref{fig:multifreq_spectra} a and e). However, the satellites are already at S-band frequencies significantly broadened by site-to-site disorder resulting in a distribution of $^{29}$Si HFIs (A-strain), which hampers the precise determination of $^{29}$Si HFIs.\\
With increasing resonance frequency, the central line and the satellites exhibit increasing asymmetric line broadening, which may be attributed to g-anisotropy and g-strain. Therefore, the satellites, which are still resolved at X-band frequencies, overlap with the central line at Q- and W-band. Since the resolution of the principal g-values requires high frequencies, it becomes impossible to extract the magnetic parameters at one single frequency. It is important to note that a complete resolution of the principal g-values is not possible even at very high frequencies (see Fig.~\ref{fig:multifreq_spectra} d and h) since g-strain is proportional to the resonance frequency.

\subsection{Modeling of multifrequency EPR spectra\label{sec:modeling_EPRspec}}
The EPR spectrum may be described by the following spin Hamiltonian ($\mathcal{H}$) of a single electron spin ($S = 1/2$) coupled to $n$ surrounding nuclei (indexed by $j$) \cite{atherton1993_book}:

\begin{equation}
\mathcal{H}=\mu_\text{B}\boldsymbol{B}_0\tensorss{g}\boldsymbol{S}/\hbar+\sum_j\mu_\text{N}g_{j\text{n}}\boldsymbol{B}_0\boldsymbol{I}_j/\hbar+\sum_j\boldsymbol{S}\tensorss{A}_j\boldsymbol{I}_j
\label{eq:native_eq1}
\end{equation}

where $\mu_\text{B}$ is the Bohr magneton and $\mu_\text{N}$ the nuclear magneton. The first term denotes the electron Zeeman interaction, which couples the electron spin $S$ to the external magnetic field, $\boldsymbol{B}_0$, via the anisotropic g-tensor, $\tensorss{g}$. The second term represents the nuclear Zeeman interaction of the coupled nuclear spins $\boldsymbol{I}_j$ with $\boldsymbol{B}_0$, where $g_{j\text{n}}$ is the isotope dependent nuclear g-factor. The third term denotes the HFI, which describes the coupling of electron and nuclear spins by the A-tensor, $\tensorss{A}$. The A-tensor can be split into its isotropic part, $\tensorss{A}_\text{iso}$ and its traceless, anisotropic part, $\tensorss{A}_\text{aniso}$. The isotropic part is given by the unit matrix times $A_\text{iso}$, which is proportional to the spin density at the nucleus (Fermi-contact interaction). In case of axial symmetry, the anisotropic part can be expressed as

\begin{equation} 
\tensorss{A}_\text{aniso} = 
\left(\begin{matrix} -A_\text{dip}&&\\&-A_\text{dip}&\\&&2A_\text{dip}
\end{matrix}\right).
\end{equation}

The g-tensor, $\tensorss{g}$, and the hyperfine tensor, $\tensorss{A}$, are 3x3 matrices with the principal values ($g_{x}$, $g_{y}$, $g_{z}$) and ($A_{x}$, $A_{y}$, $A_{z}$), respectively. Their respective principal axes systems are not necessarily collinear.\\
Due to the presence of strong g- and A-strain, the dominating A- and g-values cannot be extracted from the EPR spectra directly. Instead simultaneous simulations of the EPR spectra based on Eq.~\ref{eq:native_eq1} have to be performed. If we assume that the spins are homogeneously distributed in the material (no clustering), only one electron spin needs to be included in the calculations since in that case the spin system is sufficiently dilute \cite{atherton1993_book}. The Zeeman term in Eq.~\ref{eq:native_eq1} can then be solved exactly with the three principal values of the g-tensor as fit parameters. The magnetic-field dependent broadening induced by g-strain is explicitly included in the simulation as an uncorrelated gaussian distribution of the principal values. This procedure has the advantage that the distribution parameters can be extracted directly from the fitting routine and are therefore separated from magnetic-field independent broadening. In contrast to the Zeeman term, the treatment of the HFI term in Eq.~\ref{eq:native_eq1} is more complicated, because an exact simulation including all nuclei is impossible. We therefore introduce a fitting model with certain approximations. The HFI term in Eq.~\ref{eq:native_eq1} is usually divided into two terms
 
\begin{equation}
\mathcal{H}_\text{HFI}=\sum_{k=1}^n\boldsymbol{S}\tensorss{A}_{k}\boldsymbol{I}_k+\sum_{j\neq k}\boldsymbol{S}\tensorss{A}_j\boldsymbol{I}_j\label{eq:native_eq2}
\end{equation}

where the first term describes the resolved HFI, for which the EPR resonance positions are calculated explicitly in the simulation. The second term contains the unresolved HFIs which lead to a broadening of the magnetic resonance line. The line shape induced by unresolved HFIs is described by an empirical broadening function. This is a very convenient procedure, since the first term involves only a few nuclei, while the second term runs over a very large number of nuclei. The shape of the broadening function is usually well-described by one- or two-parameter functions like a Gaussian, Lorentzian or Voigtian \cite{weil1999_mpr,vanvleck1948_pr,kittel1953_pr} which can be fitted to the resonance line by parameter adjustment. In addition to unresolved HFIs, paramagnetic centers in the solid-state experience additional line broadening due to electron-electron spin-spin interaction and life-time broadening due to $T_1$ and $T_2$ mechanisms. However, for the low defect concentrations and the low temperatures used in this study both of these mechanisms contribute less than 1 $\mu$T to the line width and can therefore be neglected \cite{wyard1965_prs,stutzmann1983_prb}. This approximation is supported by the experimental observation that the line width of the broadening function is directly proportional to the $^{29}$Si content of a-Si:H \cite{umeda1999_prb}. We can therefore conclude that the broadening function is dominated by unresolved HFIs. It is important to note that due to the low natural abundance of $^{29}$Si ($p = 4.68$~\%) the central line portion of the broadening function exhibits a Lorentzian and not a Gaussian shape \cite{umeda1999_prb,kittel1953_pr}.\\
As in earlier studies we use a fitting model, where only one $^{29}$Si
nuclear spin is treated explicitly ($n = 1$ in Eq.~\ref{eq:native_eq2},
$\tensorss{A}_1=\tensorss{A}_\text{L}$). The HFIs of all other spin carrying nuclei (such as $^1$H, $^{29}$Si) are assumed to be unresolved and are taken into account by a Voigtian line broadening function \cite{stutzmann1989_prb,umeda1999_prb}. In addition, we limit the number of fitting parameters by introducing several prior assumptions for the symmetry of the A$_\text{L}$-tensor and the orientation between the g- and A$_\text{L}$-tensor. Since the satellites are strongly affected by inhomogeneous broadening it is difficult to test the symmetry of the A$_\text{L}$-tensor against rhombicity. Furthermore, the relative orientation between the A$_\text{L}$- and g-tensor and its distribution cannot be determined independently, so we simply assume that both tensors are collinear with $g_z$ and $A_z$ being parallel. 
The principal values of $\tensorss{A}_\text{L}$ are distributed (A-strain) and we included this effect in the simulation as an uncorrelated gaussian distribution of the principal values. It is assumed that the principal values of $\tensorss{A}_\text{L}$ are not correlated to the principal values of $\tensorss{g}$ (uncorrelated g- and A-strain).

\subsection{Multifrequency fitting algorithm}
To extract the A- and g-values we applied the following step-wise fitting routine. In a first step, the Q- and W-band spectra were fitted simultaneously by adjusting the distribution parameters of the three principal g-values (mean value and standard deviation). In a second step, the S- and X-band spectra were fitted by adjusting the distribution parameters of the A$_\text{L}$-tensor principal values, where we again assumed independent normal distributions. In a third step, the S-band spectrum was fitted by adjusting a convolutional Voigtian line broadening function accounting for inhomogeneous broadening by unresolved HFI. 
The three steps were repeated in a loop until convergence is reached. The simulations of the individual solid-state EPR spectra were performed with EasySpin, a MATLAB (The Mathworks, Natick, MA, USA) toolbox \cite{stoll2006_jmr}. Powder EPR spectra are evaluated by considering a large set of different orientations uniformly distributed over the unit sphere. The simulated spectra are fitted to the experimental spectra by nonlinear least-squares methods using a trust-region-reflective algorithm implemented in MATLAB \cite{coleman1996_siam,coleman1994_mp}. Standard errors of fit parameters indicating a significance level of $1\sigma$ are calculated by a linear sum of statistical and systematic errors. Statistical errors due to spectral noise are estimated by calculating asymptotic confidence intervals in a fixed-regressor model (for details see Ref.~\cite{seber1989_book}, chapter 5). Systematic errors arise due to imprecise measurement of the regressors (magnetic field and microwave frequency). While measurement errors of the microwave frequency are usually < 1 kHz and therefore negligible, the magnetic field at the sample position is not measured directly, but has to be calibrated with a field standard sample (LiLiF, BDPA). Typical errors of such calibrations and drifts of the magnetic field over time are 0.1 mT at high magnetic fields (Q- and W-band) and 0.02 mT at low magnetic fields (S- and X-band). We roughly estimated the impact of these measurement errors on the fit parameters by repeating the above multifrequency fit routine for a worst case scenario, where all magnetic-field axes are offset by the estimated measurement error. The obtained errors of the fit parameters are then assigned to a standard error (significance level $1\sigma$) to indicate the uncertainty of the fit parameter values.

\subsection{Multifrequency fit results}
The fit results for a rhombic g-tensor are shown as solid lines in Fig.~\ref{fig:multifreq_spectra} a-d. In earlier publications it was explicitly assumed that the g-tensor is axially symmetric, i.e. $g_x=g_y=g_\perp$~and~$g_z=g_\parallel$. In order to test this hypothesis, we performed two separate multifrequency fits. In a first fit the symmetry of the g-tensor is forced to axial symmetry and in a second fit (see Fig.~\ref{fig:multifreq_spectra} a-d) no assumptions about the symmetry were made. In the first case the principal values of the g-tensor are $g_x=g_y=2.0065(2)$ and $g_z=2.0042(2)$, in very good agreement with earlier studies (see Ref.~\cite{umeda1999_prb} and Table~\ref{tab:multifreq_study}). In the second case we obtained a rhombic g-tensor with three different principal values ($g_x=2.0079(2)$, $g_y=2.0061(2)$ and $g_z=2.0034(2)$). 
However, the quality of the fit, measured by the sum of squares of the fit residuals $\left\Vert r \right\Vert_2^2$ (difference between the fitted and the experimental spectra), is significantly worse in the case of an axially symmetric g-tensor as compared to a rhombic g-tensor (see Table~\ref{tab:multifreq_study}).
On the basis of our fit results, we can state that $g_x$ and $g_y$ do not coincide on a significance level of $\approx 5\sigma$. We therefore conclude that coordination defects in a-Si:H exhibit a rhombic g-tensor.\\

For the A$_\text{L}$-tensor we obtained $A_x=A_y=151(13)$ MHz and $A_z=269(21)$ MHz, which corresponds to $A_\text{iso}=190(11)$ MHz and $A_\text{dip}=39(8)$ MHz. Please note that we only report magnitude values for the HFIs since FSE-EPR does not provide the sign of the HFIs. These values are slightly smaller than the previously reported ones (see Table \ref{tab:multifreq_study}). The Voigtian broadening function, accounting for unresolved HFI, deviates only slightly from a pure Lorentzian function since the FWHM of the Gaussian component is about a factor of 4 smaller than the FWHM of the Lorentzian component (see Table \ref{tab:multifreq_study}).
A complete overview of the various fit parameter sets including literature values is given in Table \ref{tab:multifreq_study}.

\subsection{DFT calculations of DB g- and A-tensors}
\begin{figure}
\includegraphics{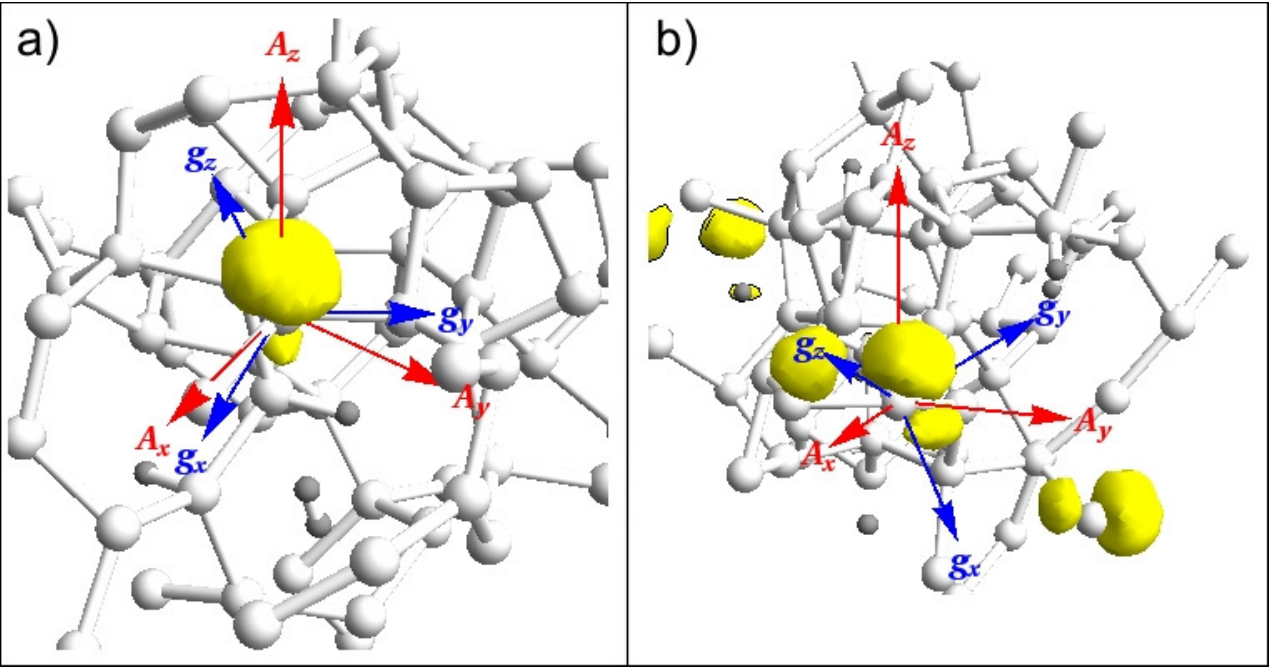}
\caption{Two selected computer-generated atomistic models of a DB in a-Si:H (for details see text). Principal values of $\tensorss{g}$ and $\tensorss{A}_\text{L}$ for the two models are a) $g_x$ = 2.0091, $g_y$ = 2.0057, $g_z$ = 2.0024, and $A_x$ = -291 MHz, $A_y$ = -288 MHz, $A_z$ = -427 MHz, b) $g_x$ = 2.0095, $g_y$ = 2.0065, $g_z$ = 2.0034, and $A_x$ = -176 MHz, $A_y$ = -180 MHz, $A_z$ = -236 MHz . Atoms are indicated by light-gray (Si) and dark-gray (H) shaded spheres. Isosurface plot of the electron spin density (isosurface at 10~\% of maximum spin-density value) of the trivalent silicon atom is show in yellow. Eigenvectors of the g-tensor are indicated by blue arrows and the eigenvectors of the A$_\text{L}$-tensor of the threefold-coordinated Si atom are indicated by red arrows.
\label{fig:multifreq_spin_plot}}
\end{figure}

The above analysis of the experimental spectra provided g- and A-values of paramagnetic coordination defects present in a-Si:H. Our task is now to deduce the microscopic origin of the defect centers from the obtained interaction values. In order to test the hypothesis that coordination defects in a-Si:H are DB defects we examined g- and A-values of computer-generated DB defect models by DFT calculations. As outlined above, DB defects were created in a-Si:H computer models by removing a single H atom from defect-free structures. However, it is important to note that theoretical modeling of DB defects in a-Si:H is a demanding task since the atomic defect structure is not well defined as in case of $P_b$ defects at the Si/SiO$_2$ interface. Disorder in amorphous materials induces a large variety of atomic configurations. In order to account for this variety we modeled a large number of different defect structures (26 in total) and calculated the resulting g- and A-values. The aim of this approach is to link the observed g- and A-values with particular features of the atomic structure. However, to our surprise we found that quite different spin-density distributions result in very similar g-values.\\
To illustrate this finding we compare calculated ground-state spin-densities, g- and A$_\text{L}$-tensors of two particular computer-generated DB models. The first DB model (DB1) is displayed in Fig.~\ref{fig:multifreq_spin_plot}a ($g_x$ = 2.0091, $g_y$ = 2.0057, $g_z$ = 2.0024, and $A_x$ = -291 MHz, $A_y$ = -288 MHz, $A_z$ = -427 MHz for the $^{29}$Si A$_\text{L}$-tensor) and the second DB model (DB2) is shown in Fig.~\ref{fig:multifreq_spin_plot}b ($g_x$ = 2.0095, $g_y$ = 2.0065, $g_z$ = 2.0034, and $A_x$ = -176 MHz, $A_y$ = -180 MHz, $A_z$ = -236 MHz for the $^{29}$Si A$_\text{L}$-tensor). The g-tensor symmetry of both models is clearly rhombic, while the A$_\text{L}$-tensor is very close to axial symmetry. Already from a superficial inspection of the two structures it becomes apparent that the wave function of DB1 is mainly localized on a single Si atom, while the wave function of DB2 is more delocalized. Despite the apparent discrepancy of the spin-density distributions, the g-tensor principal values of both models are almost identical. Hence, widely different configurations can yield almost identical g-tensors. This effect will be discussed in more detail below. In contrast to the almost identical g-tensor, the HFI and the relative orientation between g- and A$_\text{L}$-tensor vary drastically. The isotropic HFI of DB2 ($A_\text{iso}$ = -197 MHz) is much smaller than the isotropic HFI of DB1 ($A_\text{iso}$ = -335 MHz), which can be attributed to a delocalization of the DB wave function. The axes of the $g_z$ and $A_z$ principal values are nearly parallel in case of DB1 but differ significantly in case of DB2 (see Fig.~\ref{fig:multifreq_spin_plot}a,b).

\subsection{Comparison of experimental and theoretical results}
We have observed that the two computer-generated defect structures analyzed above show a substantial variation in terms of spin-density distribution. This variation is a result of site-to-site disorder present in a-Si:H leading to a wide distribution of A- and g-values (A- and g-strain). It is therefore clear that a comparison of g- and A-values from only one or two computer models is not sufficient for a successful identification of the microscopic origin of defect centers in a-Si:H. Instead, it is mandatory to evaluate a representative number of DB models and their electronic structure to cover the whole spread of g- and A-value distributions. We therefore extend our analysis to a larger set of defect models, which includes 26 DB models in total. Histograms of g- and A-values of those models are shown in Fig.~\ref{fig:multifreq_g} and \ref{fig:multifreq_HFI} and compared to experimental distribution functions obtained by the multifrequency fit. A compilation of principal values of $\tensorss{g}$ and $\tensorss{A}_\text{L}$ and plots of spin-density distributions for each DB model can be found in the supplemental material \cite{suppl_mat}. An inspection of the spin-density distribution of individual DB defects shows that the majority of defects exhibits a spin-density distribution which is bound to a single, undercoordinated, Si atom. In the following we will compare the g- and A-tensors obtained from DFT with the respective parameters extracted from multifrequency EPR data.

\subsubsection{g-tensor}

\begin{figure}
\includegraphics[width=0.8\textwidth]{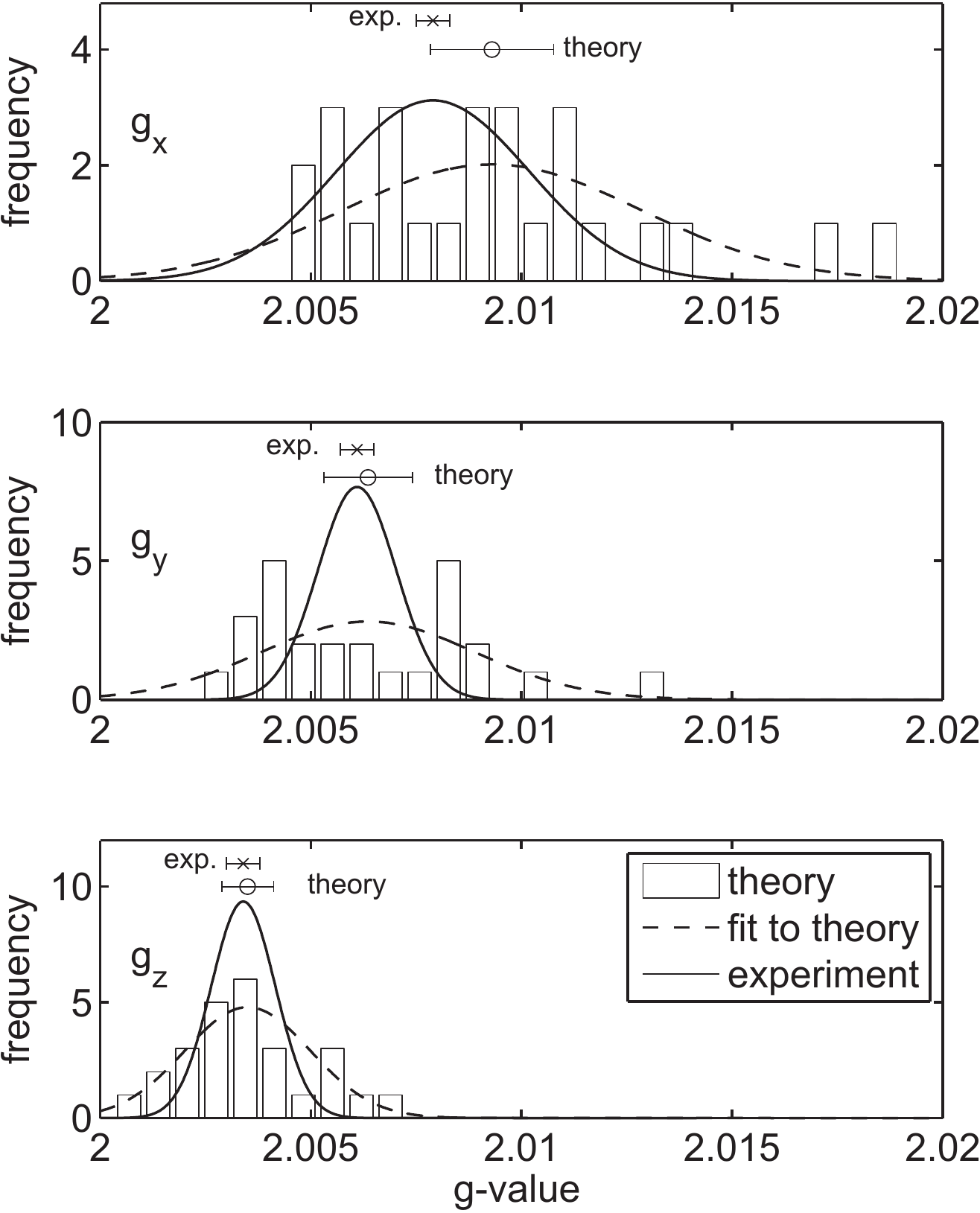}
\caption{Comparison between experimentally and theoretically (DFT) obtained principal g-values of coordination defects in a-Si:H. Values for different computer-generated DB models in a-Si:H are shown by the histogram. Principal values are sorted by size and assigned to $g_x$, $g_y$ and $g_z$. The histogram was fitted by a normal distribution function shown by the dashed line. The experimental data for coordination defects obtained by a fitting model is shown by the solid line. The confidence intervals ($2\sigma$) of the mean experimental (cross) and theoretical (circle) principal values are shown separately to indicate their statistical significance.
\label{fig:multifreq_g}}
\end{figure}

In Fig.~\ref{fig:multifreq_g} the distribution of principal values of \tensorss{g} derived from computer models are plotted together with experimental distribution functions. The distribution mean values and width of theoretically obtained values and their uncertainties can be estimated by fitting a normal distribution to the g-values. The mean of the calculated values are $g_x=2.0093(7)$, $g_y=2.0064(5)$ and $g_z=2.0035(3)$. Comparing those values to the experimental results ($g_x=2.0079(2)$, $g_y=2.0061(2)$ and $g_z=2.0034(2)$) shows that the experimental $g_y$ and $g_z$ principal values deviate less than one $\sigma$ from the theoretical values, while $g_x$ deviates about two $\sigma$ (see also error bars in Fig.~\ref{fig:multifreq_g}). This analysis shows that there is a good agreement for the $g_y$ and $g_z$ mean principal values, while there is a significant deviation in the case of $g_x$. In all three cases the spread of the computed principal g-values significantly exceeds the spread of the experimental distributions (see Fig.~\ref{fig:multifreq_g} and Table~\ref{tab:multifreq_study}).\\
By inspecting distributions of $g_x$, $g_y$ and $g_z$ separately, we see that they peak at different values, although parts of the distributions overlap. By analyzing Fig.~\ref{fig:multifreq_g}, it becomes clear that the $g_z$ distribution peaks close to the free-electron g-value ($g_e=2.0023$) and is well separated from the $g_x$ and $g_y$ distribution. However, since there is a large overlap between distributions of $g_x$ and $g_y$, it is not immediately clear whether the distributions are independent or if $g_x$ and $g_y$ actually belong to the same distribution as it would be the case for an axially symmetric g-tensor. In that case the distribution would be much wider, but still most of the g-tensors would exhibit a slight rhombic symmetry. It is therefore necessary to determine the g-tensor rhombicity of each DB model separately by calculating $(g_x-g_y)/(g_x-g_z)$. By doing so we found that each individual g-tensor is clearly rhombic and the distribution peaks at 0.5 which fits well to the experimentally obtained symmetry.\\
It appears to be surprising that g-tensors of DB defects with a very symmetric spin-density distribution (see Fig.~\ref{fig:multifreq_spin_plot}a) exhibit rhombic symmetry and not axial symmetry. This effect can be rationalized as follows. The anisotropy of the g-tensor and isotropic shifts from $g_e$ result from an indirect coupling of the electron spin to the external magnetic field mediated by the orbital momentum. In the picture of second order perturbation theory \cite{stone1963_prs}, the most important contribution arises from the interplay of the singly-occupied DB orbital $\psi_p$ with all unoccupied or occupied orbitals $\psi_n$ other than $\psi_p$ (orbital energies $\epsilon_n$ and $\epsilon_p$) weighted by their inverse energetic separation,

\begin{equation}
g_{\alpha\beta}=\delta_{\alpha\beta}g_e+...+2\sum_{n\neq p}\frac{\langle \psi_{p}|\lambda L_{\alpha}|\psi_{n}\rangle\langle \psi_{n}|L_{\beta}|\psi_{p}\rangle}{\epsilon_p-\epsilon_n}
\label{eq:stone}
\end{equation}

Here, $\boldsymbol{L}$ denotes the angular momentum operator and $\lambda$ the spin-orbit coupling constant. Even in case of a DB orbital, which is completely localized at the threefold-coordinated Si atom, the g-tensor is obviously sensitive to changes in the orientation and energies of the other orbitals. To simplify the discussion of the g-tensor anisotropy, let us assume that the singly-occupied DB orbital $\psi_p$ is given by a pure $\left|p_z\right\rangle$ orbital and the other orbitals $\left|p_{x,y}\right\rangle$ are also of atomic type. We define the Cartesian coordinate system such that the z-axis coincides with the axis of the DB orbital. By this we can show that the paramagnetic contribution vanishes for $\alpha\beta=zz$, since $L_z\left|p_z\right\rangle=0$. Significant deviations from $g_e$ are therefore only expected for $g_x\equiv g_{xx}$ and $g_y\equiv g_{yy}$ given by

\begin{eqnarray}
\Delta g_{xx}=2\frac{\langle p_z|\lambda L_{x}|p_y\rangle\langle p_y|L_x|p_z\rangle}{\epsilon_{p_z}-\epsilon_{p_y}}&=& 2\frac{\lambda}{\epsilon_{p_z}-\epsilon_{p_y}},\nonumber\\
\Delta g_{yy}=2\frac{\langle p_z|\lambda L_{y}|p_x\rangle\langle p_x|L_{y}|p_{z}\rangle}{\epsilon_{p_z}-\epsilon_{p_x}}&=& 2\frac{\lambda}{\epsilon_{p_z}-\epsilon_{p_x}}.
\label{eq:native_eq4}
\end{eqnarray}

We see that if the degeneracy of the $p_x$ and $p_y$ orbitals is lifted, the $g_{xx}$ and $g_{yy}$ values will not be degenerate. In a most disordered environment like a-Si:H one expects that the degeneracy is lifted due to fluctuations of the bond-angles and bond-length. As a result a rhombic g-tensor instead of an axially symmetric one arises. This analysis is also valid for the more realistic case of $\psi_{p,n}$ being molecular orbitals.\\
We have seen that there is a quantitatively good agreement of the calculated g-tensors of DB models and the experimentally determined g-tensor of coordination defects in a-Si:H. However, we have shown that the g-tensor principal values are rather insensitive to the spin-density distribution of DBs. Widely different wave functions give rise to almost identical g-tensors. We now extend our analysis to the principal values of $\tensorss{A}_\text{L}$, which are a more precise probe of the local spin-density distribution of the defect center.

\subsubsection{Hyperfine interactions}
\begin{figure}
\includegraphics[width=\textwidth]{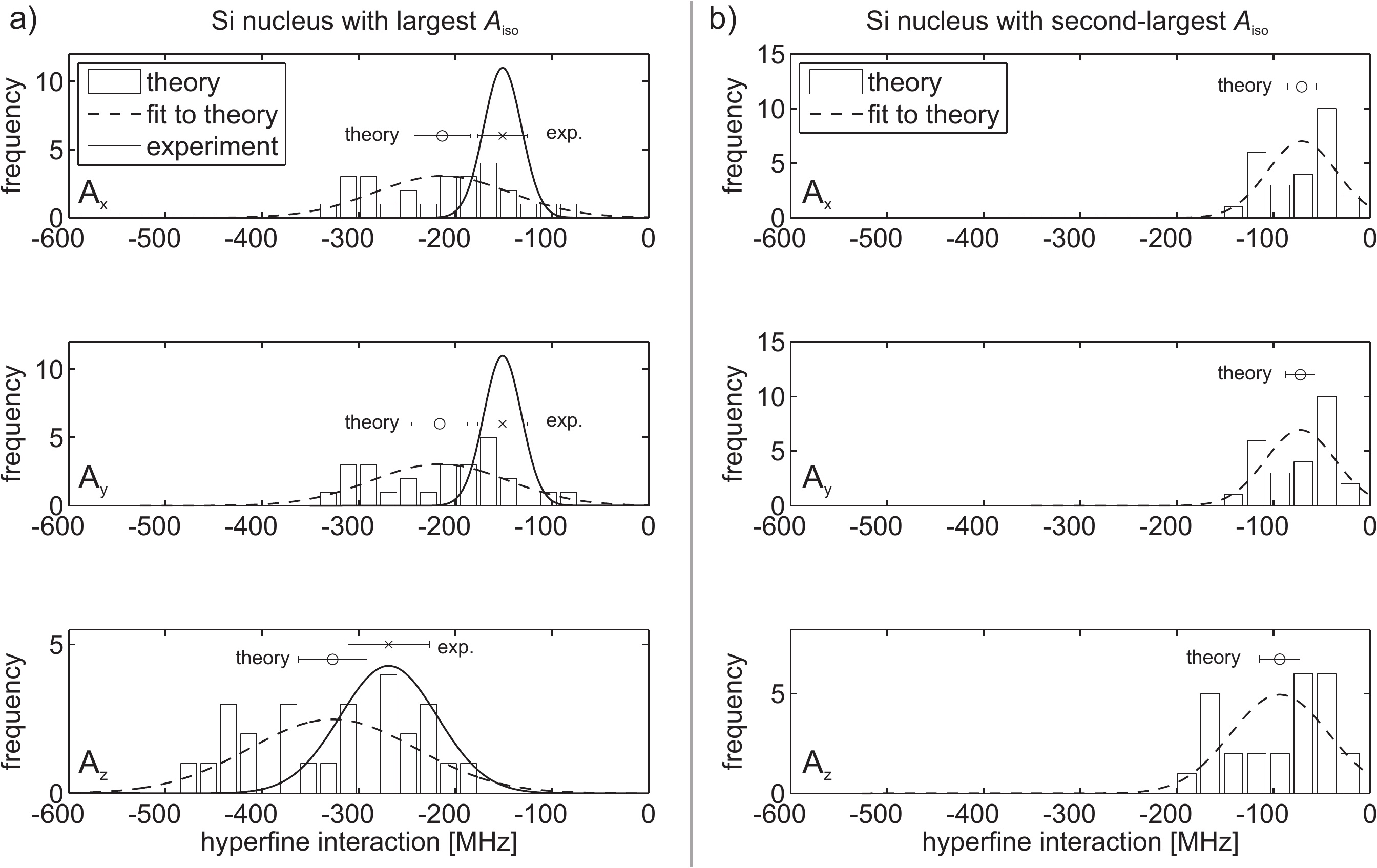}
\caption{Comparison between experimentally and theoretically (DFT) obtained principal values of the largest and second-largest HFI of coordination defects in a-Si:H. A-tensor of the $^{29}$Si nucleus with the largest $A_\text{iso}$ (a) and second largest $A_\text{iso}$ (b). Values for different computer-generated models of a DB in a-Si:H are shown by the histogram. Principal values are sorted by size and assigned to $A_x$, $A_y$ and $A_z$. The histogram was fitted by a normal distribution function shown by the dashed line. The experimental data for coordination defects obtained by a fitting model is shown by the solid line. The confidence intervals ($2\sigma$) of the mean experimental (cross) and theoretical (circle) principal values are shown separately to indicate their statistical significance.
\label{fig:multifreq_HFI}}
\end{figure}

The histogram in Fig.~\ref{fig:multifreq_HFI}a shows a comparison of principal A$_\text{L}$-values derived from different computer-generated DB models with experimentally obtained values. Theoretically obtained distributions were approximated by normal distributions. It is found that values derived from computer-generated DBs deviate from values determined by the multifrequency fit.
The absolute mean of all three principal values of $\tensorss{A}_\text{L}$ obtained by theory are $A_x = 213(14)$ MHz, $A_y = 216(14)$ MHz and $A_z = 327(17)$ MHz, whereas the experimental values are $A_x = 151(13)$ MHz, $A_y = 151(13)$ MHz and $A_z = 269(21)$ MHz. The theoretical HFI values are therefore larger than the experimental values by at least $2\sigma$ (see also error bars in Fig.~\ref{fig:multifreq_HFI}a)). The discrepancy of $\tensorss{A}_\text{L}$ between experiment and theory is therefore much more pronounced than in the case of $\tensorss{g}$. As observed for the g-tensor, the calculations lead to a larger distribution of principal values as compared to experiment.\\
$A_x$ and $A_y$ distributions in computer models are clearly degenerate and strongly differ from $A_z$. This hints towards an A-tensor close to axial symmetry. To check whether the $A_x$ and $A_y$ distributions of the computer models are independent or not, we determined the rhombicity $(A_x-A_y)/(A_x-A_z)$ of each set of principal values and found that the A$_\text{L}$-tensors for all models investigated are indeed very close to axial symmetry. These results support the previously made assumption of an axially symmetric A$_\text{L}$-tensor for the fits of the EPR spectra in Fig.~\ref{fig:multifreq_spectra}.\\
The calculations revealed a rather peculiar deviation of the symmetry properties of $\tensorss{g}$ and $\tensorss{A}_\text{L}$, where $\tensorss{g}$ exhibits rhombic symmetry and $\tensorss{A}_\text{L}$ axial symmetry. The apparent discrepancy in the symmetry properties can be rationalized as follows. The A$_\text{L}$-tensor depends directly on the ground-state spin density and is strongly dominated by the local-orbital character (sp$^x$ hybrid) of the DB state at the site of the trivalent Si atom. Structural variations due to the amorphous matrix affect its orientation and possibly the degree of s-p hybridization, but do not alter the fundamental sp$^x$ character of the DB orbital. Its axial symmetry properties are therefore maintained even in the presence of large disorder-induced fluctuations of the bond-length and bond-angles in a-Si:H. This does also hold in good approximation for the HFIs of $^{29}$Si atoms in the first and second coordination shell.\\
We have seen that some of our defect models exhibit a significant spin delocalization. Yet, all defect models exhibit typical DB characteristics of a spin density mainly localized on a single, undercoordinated atom (referred to as the central atom), while the spin density on the other atoms is significantly smaller. This is reflected by the fact that Si atoms with the second-largest HFI have values that average to $A_\text{iso} = -79(5)$ MHz and $A_\text{dip} = -7(4)$ MHz (see Fig.~\ref{fig:multifreq_HFI}b).\\
We have seen that the mean principal values of $\tensorss{A}_\text{L}$ deviate between theory and experiment. Decomposing $\tensorss{A}_\text{L}$ into an isotropic and anisotropic part clearly shows that this discrepancy arises from the isotropic part ($A_\text{iso}$) while the anisotropic HFI ($A_\text{dip}$) in both cases equals about 35 to 40 MHz (see Table~\ref{tab:multifreq_study}). The most puzzling fact comparing computed and experimentally obtained HFIs is therefore the discrepancy of the mean isotropic HFI with the following values for theory and experiment:

\begin{eqnarray}
\text{theory}:&\left|A_\text{iso}\right| &= 252(9)~\text{MHz}\\
\text{experiment}:&\left|A_\text{iso}\right| &= 190(11)~\text{MHz}.
\end{eqnarray}

These values reveal two important findings. Firstly the experimentally obtained $A_\text{iso}$ of coordination defects in a-Si:H is smaller than the value derived from the DFT calculations with a level of confidence of more than $2\sigma$.
Secondly, both values are much smaller than $A_\text{iso}$ of $P_b$ centers at the Si/SiO$_2$ interface (315~MHz)~\cite{brower1983_apl,carlos1987_apl}, which was frequently employed as model system for the coordination defects in a-Si:H.
The first finding may be rationalized by an inspection of the computer-generated spin density maps. We find that the lower value of $A_\text{iso}$ in a-Si:H as compared to $P_b$ centers is primarily caused by a delocalization of the DB spin density. 
In contrast to previous assumptions, we did not find any evidence for a relaxation of the atomic structure which could lead to a reduction of $A_\text{iso}$~\cite{stutzmann1989_prb}. 
Our results render the relaxation of DBs in a-Si:H towards a more planar defect geometry improbable.
Hence, the observed deviation between experimental and theoretical values of $A_\text{iso}$, our second finding, must be of different origin. One possible explanation is that the chosen population of defect structures in the DFT calculation may not represent the paramagnetic site in a-Si:H. This again raises a heavily debated question: Which kind of coordination defect gives rise to the EPR signal centered around $g = 2.0055$? Due to the limited number of model structures in this work, it cannot be excluded that coordination defects in a-Si:H form DB structures, which are not contained in the DFT defect pool or that the microscopic structure of the defects completely differs from a DB. If coordination defects exhibit more delocalized spin densities than random DBs considered in this work, the largest isotropic HFI will be smaller. A more delocalized defect structure is therefore one possibility to explain the discrepancy between experiment and model calculations.\\
In view of these results, it is evident that the structural models employed for the theoretical analysis miss an important aspect of the experimentally observed defect ensemble. An obvious weakness of the theoretical modeling is that the DBs were created at random points in the amorphous network and were subject only to local relaxation. More complex, but slow ($>10$ ns) relaxation mechanisms possibly occurring in the real material are therefore not captured at all. If present, such relaxations might select a subset of the present defect models, or even other configurations. For instance, the floating-bond type defect exhibits states delocalized over several Si atoms \cite{fedders1988_prb,biswas1989_prl}. However, the floating-bond model has long been rejected, being in conflict with a number of other experimental observations \cite{fedders1988_prb,biswas1989_prl,fedders1989_prb_nonideal,ishii1990_prb}. At present we can only speculate over plausible microscopic defect models since the available data do not allow us to discriminate them.

\section{Conclusions}
Using a multifrequency approach, we have determined g-tensor principal values of coordination defects in a-Si:H: $g_x=2.0079$, $g_y=2.0061$ and $g_z=2.0034$ with improved accuracy (see Table~\ref{tab:multifreq_study}). In contrast to earlier studies \cite{stutzmann1989_prb,umeda1999_prb}, we found that the g-tensor shows pronounced rhombicity. 
In addition, we carried out a first systematic study where experimental g- and A$_\text{L}$-values ($^{29}$Si-HFI tensor with the largest isotropic part) are compared to theoretical values obtained by DFT calculations of 26 different a-Si:H DB models. As main conclusions we found that computer models reproduce the experimentally observed principal values and rhombicity of the g-tensor, but do not exhibit HFIs in agreement with experiment. The apparent discrepancy between symmetry properties of g- and A-tensors is attributed to the fact that the g-tensor reflects the global electronic defect structure while the A-tensor is exclusively determined by the local spin-density distribution in the vicinity of the nucleus of interest. This leads to a situation where DBs with a localized and a delocalized spin-density distribution exhibit almost identical g-tensors. Principal values of the A$_\text{L}$-tensor for computer generated DB models disagree with experimental values obtained by the multifrequency fit. The isotropic HFI of the DB models is on average $A_\text{iso}=-252$ MHz which is much larger than the fit result, $A_\text{iso}=190$ MHz. Our DFT calculations do not support the hypothesis formulated in earlier studies that the structure of DB defects relaxes towards a more planar geometry and thereby reduces the isotropic HFI. These observations strongly suggest that coordination defects in a-Si:H are more delocalized than investigated DB computer models. We therefore conclude that coordination defects in a-Si:H are not well-described within the random DB model. However, to develop plausible alternative models, additional DFT studies are required. Such studies are on the way within the research network \textit{EPR-Solar}.

\begin{acknowledgments}
Financial support from BMBF (EPR-Solar network project 03SF0328) is acknowledged. We are very grateful to F. Inam (ICTP, Italy) and D. Drabold (Ohio University, USA) as well as K. Jarolimek (TU Delft, Netherlands) for providing us the DB model structures used in the theoretical calculations.  U. Gerstmann (University of Paderborn, Germany), M. Stutzmann and M. Brandt (TU Munich, Germany) are gratefully acknowledged for helpful discussions.
\end{acknowledgments}


%

\begin{table*}
\scriptsize
\caption{\label{tab:multifreq_study}Summary of experimental (multifrequency fit) and theoretical g-tensor and A$_\text{L}$-tensor principal values for coordination defects in a-Si:H. Full-width half maximum (FWHM) of gaussian distributions of the g- and A$_\text{L}$-tensor principal values (g- and A-strain) are given in square brackets. The Voigt function accounting for magnetic-field independent broadening is characterized by FWHM of Gaussian and Lorentzian components denoted by $\Delta B^\text{G/L}$. Standard errors of the fit parameters (significance level $1\sigma$) are calculated by a linear sum of statistical and systematic errors (for details see text) and are given in round brackets. $\left\Vert r \right\Vert_2^2$ denotes the sum of squares of the fit residual $r$ in units of the sum of squares of the experimental data.}
\begin{ruledtabular}
\begin{tabular}{lcccccccp{2.5cm}c}
&\multicolumn{3}{c}{principal values of g-tensor}&\multicolumn{4}{c}{principal values of $^{29}$Si A$_\text{L}$-tensor}&broadening function&\\ \hline
&$g_x$ or $g_\perp$&$g_y$ or $g_\perp$&$g_z$ or $g_\parallel$& $A_x$ or $A_\perp$&$A_y$ or $A_\perp$&$A_z$ or $A_\parallel$&$A_\text{iso}$/$A_\text{dip}$&Voigtian&$\left\Vert r \right\Vert_2^2$\\
&[strain]\footnotemark[1]&[strain]\footnotemark[1]&[strain]\footnotemark[1]&[strain]&[strain]&[strain]&&$\Delta B^\text{G/L}$&\\
&&&&in MHz\footnotemark[2]&in MHz\footnotemark[2]&in MHz\footnotemark[2]&in MHz\footnotemark[2]&in mT&in \%\\ \hline
(EPR)&&&&&&&&\\
Present\footnotemark[3]&2.0079(2)&2.0061(2)&2.0034(2)&151(13)&151(13)&269(21)&190(11)/39(8)&0.13(3)/0.43(1)&0.2\\
&[0.0054(1)]&[0.0022(1)]&[0.0018(1)]&[46(27)]&[46(27)]&[118(66)]&&&\\
Present\footnotemark[4]	&2.0065(2)&2.0065(2)&2.0042(2)&149(15)&149(15)&265(26)&188(13)/39(10)&0.15(3)/0.42(1)&0.3\\
												&[0.0047(1)]&[0.0047(1)]&[0.0019(1)]&[47(32)]&[47(32)]&[113(75)]&&&\\[40pt]
Ref. \cite{umeda1999_prb}&2.0065&2.0065&2.0039&143&143&333&206/63&N/A\footnotemark[5]\\
&[N/A]\footnotemark[5]&[N/A]\footnotemark[5]&[N/A]\footnotemark[5]&[56]&[56]&[73]&&\\
Ref. \cite{stutzmann1989_prb}&2.0080&2.0080&2.0040&154&154&305&205/50&not specified\\
&[0.0029]&[0.0029]&[0.0022]&[28]&[28]&[56]&&\\[40pt]
(Theory)&&&&&&&&\\
Present\footnotemark[6]&2.0093(7)&2.0064(5)&2.0035(3)&-213(14)&-216(14)&-327(17)&-252(9)/-37(7)&non-analytic\footnotemark[7]\\
&[0.0084(12)]&[0.0060(9)]&[0.0035(5)]&[166(24)]&[166(24)]&[203(30)]&&\\
\end{tabular}
\end{ruledtabular}
\footnotetext[1]{$\Delta B$ value given in mT converted to $\Delta g$
(dimensionless) using $\Delta g=\left(g_e^2\mu_\text{B}/h\right)\cdot\left(\Delta B/\nu\right)$}
\footnotetext[2]{Hyperfine interactions given in mT converted to MHz using $\nu_\text{HFI}=g_e\mu_{\text{B}}B/h$}
\footnotetext[3]{Multifrequency fit without prior assumptions about the symmetry of $\tensorss{g}$}
\footnotetext[4]{Multifrequency fit assuming axial symmetry of $\tensorss{g}$}
\footnotetext[5]{g-strain and magnetic-field independent broadening are entangled in the analysis of Ref.~\cite{umeda1999_prb} and could not be separated.}
\footnotetext[6]{DFT calculation of DB defect center in a relaxed a-Si$_{64}$H$_{7}$ supercell}
\footnotetext[7]{The broadening function of the DFT data cannot be expressed in closed analytic form.}
\end{table*}


%

\end{document}